\documentclass[journal]{IEEEtran}
\usepackage{amsfonts}
\usepackage{cite}
\usepackage{graphicx}
\usepackage{amsmath}
\usepackage{amssymb}
\usepackage{algorithm}
\usepackage{algorithmic}

\begin{document}

\title{Joint Network and LDPC Coding for Bi-directional Relaying}

\author{Xiaofu~Wu, ~Weijun~Zeng,
        ~Chunming~Zhao,
        and Xiaohu~You 
\thanks{This work was supported in part by the National High Technology Research
        and Development Program (863 program) of China under Grants 2009AA01Z235, 2008AA12Z307, and  by the National Science
        Foundation of China under Grants 60972060, 61032004. The work of X. Wu was also supported by the National Key S\&T Project under Grant 2010ZX03003-003-01 and by the Open Research Fund of National Mobile Communications Research Laboratory, Southeast University (No. 2010D03).
        }
\thanks{Xiaofu Wu is with the Nanjing Institute of Communications
        Engineering, PLA Univ. of Sci.\&Tech., Nanjing 210007, China. He is also with the National Mobile Commun. Research Lab., Southeast Univ., Nanjing
        210096 (Email: xfuwu@ieee.org).}
\thanks{Weijun Zeng is with the PLA Univ. of Sci.\&Tech., Nanjing 210007, China (Email:
        zwj3103@126.com).}
\thanks{Chunming Zhao, and Xiaohu You are with the National Mobile Commun. Research Lab.,
        Southeast University, Nanjing 210096, China (Email: cmzhao@seu.edu.cn, xhyu@seu.edu.cn).}}


\maketitle

\begin{abstract}
In this paper, we consider joint network and LDPC coding for practically implementing the denosie-and-forward protocol over bi-directional relaying.
the closed-form expressions for computing the log-likelihood ratios of the network-coded codewords have been derived for both real and complex multiple-access channels.
It is revealed that the equivalent channel observed at the relay is an asymmetrical channel, where the channel input is the XOR form of the two source nodes.
\end{abstract}

\begin{keywords}
bi-directional relaying, network coding, denoise-and-forward, LDPC coding.
\end{keywords}

\IEEEpeerreviewmaketitle

\section{Introduction}
\PARstart{N}{etwork} coding has shown its power for disseminating information over networks \cite{KoetterAlg,Chou}.
For wireless cooperative networks, there are increased interests in employing the idea of network
coding for improving the throughput of the network.
Indeed, the gain is very impressive for the special bi-directional relaying
scenarios with two-way or multi-way traffic as addressed in \cite{Popovski}.

The denoise-and-forward protocol has shows its excellence for the problem of two transmitters wishing to exchange
information through a relay in the middle. Although there are various works addressing the denoise-and-forward protocal \cite{PopovskiVTC,Popovski,KoikeJSAC},
it is still a hot topic for how to implement it in practice when the channel coding is involved. In \cite{AZhan_NNSP08,SLZhang_JSAC}, joint network and channel coding was proposed for the simple real additive multiple-access white Gaussian noise channel. By noticing the linearity of both network and channel coding, the soft Log-likelihood Ratios (LLRs) for the network-coded codeword can be directly estimated from the received physically-superimposed signals.

In this paper, we provide further insights into the joint network and LDPC coding approach for bi-directional relaying with BPSK signaling.
In particular, the complex multiple-access channel is considered and a closed-form expression is derived for the LLRs of network-coded codeword.
It is also revealed that the equivalent channel observed at the relay is an asymmetrical channel.

\section{Bi-directional Relaying  with Denoise-and-Forward}
For bi-directional relaying, we assume that communication takes place in two phases - a multiple
access (MAC) phase and a broadcast phase.

\subsection{Real Multiple Access Channel Model}

During the MAC phase, the source nodes A and B transmit the modulated signals $x_a$ and $x_b$ to the relay.
Under a real multiple-access white Gaussian noise channel, the received signal at the relay for the $k$th time epoch can be
expressed as
\begin{equation}
 \label{eq:1}
    y_r(k)=x_a(k)+x_b(k)+w_r(k),
\end{equation}
where $w_r(k)$ is an additive white Gaussian noise with zero mean and variance of $\sigma^2$.

For coded transmission, it is of interest to consider a block of transmission. In this paper,
the block coded BPSK transmission is assumed and the same codeword length $N$ is assumed for both nodes A and B.
At both source nodes, the information bits are first input to the block encoder and the encoder output the encoded vector $\mathbf{c}_a = (c_a(0),c_a(1),\cdots, c_a(N-1))^T$ for node A and $\mathbf{c}_b = (c_b(0),c_b(1),\cdots, c_b(N-1))^T$ for node B.
The encoded vector is further mapped to $\mathbf{x}=(x(0),x(1),\cdots,x(N-1))^T$ by $x(k)=2c(k)-1$ before transmission for both nodes A and B.
At the relay node, we get the received vector $\mathbf{y}_r=(y_r(0),y_r(1),\cdots,y_r(N-1))^T$, which can be formulated as
 \begin{equation}
 \label{eq:2}
    \mathbf{y}_r = \mathbf{x}_a + \mathbf{x}_b + \mathbf{w}_r.
\end{equation}

\subsection{Denoise-and-Forward}
The denoise-and-forward approach is first proposed in \cite{PopovskiVTC}. For this approach, the relay employs a denoising function based
on an adaptive network coding to map the received signal vector into a quantized signal vector $\mathbf{x}_r$ for the broadcast phase.
In theory, this is rather involved and the joint maximum likelihood sequence estimation (MLSE) is often employed. In essence,
the joint MLSE  tries to find
 \begin{equation}
 \label{eq:3}
    (\hat{\mathbf{x}}_a, \hat{\mathbf{x}_b}) = \min_{\mathbf{x}_a \in C_a, \mathbf{x}_b \in C_b} |\mathbf{y}_r - \mathbf{x}_a - \mathbf{x}_b|,
\end{equation}
It is clear that the triple $(\hat{\mathbf{x}}_a, \hat{\mathbf{x}_b}, \mathbf{y}_r)$ is jointly typical \cite{Popovski}. Then, the relay often employs a network coding
function for denoising from $\mathbf{y}_r$ to $\mathbf{x}_r$:
 \begin{equation}
 \label{eq:4}
    \mathbf{x}_r = \Upsilon (\hat{\mathbf{x}}_a, \hat{\mathbf{x}}_b).
\end{equation}
Often, the simple XOR function is enough, namely,  $\mathbf{x}_r = \hat{\mathbf{x}}_a \oplus \hat{\mathbf{x}}_b$.

During the broadcast phase, the relay transmits the signal vector $\mathbf{x}_r$ to both nodes A and B. Hence, nodes A and B can retrieve their own information  as they know completely what they have sent.

In practice, the joint MLSE is often infeasible due to the complexity issue. Hence, one often simply considers the un-coded case \cite{KoikeJSAC}, which, however, is far away from the optimality.

\section{Joint Coding over Real MAC Channel}
In this paper, we restrict the block coding scheme on LDPC coding for both nodes A and B.
Let $C_a$ be a $(N, K_a)$ LDPC code of block length $N$ and dimension
$K_a$ for node A, which has a parity-check matrix $H_a=[h_{m,n}]$ of $M$ rows, and
$N$ columns. Let $R_a=K_a/N$ denote its code rate.
Correspondingly, we can define the code $C_b$ with a parity-check matrix of $H_b$ for node B.

For any given LDPC encoded vector $\mathbf{c}_a = (c_a(0),c_a(1),\cdots, c_a(N-1))^T$ for node A and
$\mathbf{c}_b = (c_b(0),c_b(1),\cdots, c_b(N-1))^T$ for node B, we have
\begin{eqnarray}
 \label{eq:7}
    H_a \mathbf{c}_a &=& \mathbf{0}, \nonumber \\
    H_b \mathbf{c}_b &=& \mathbf{0}.
\end{eqnarray}

\subsection{Single-User Approach}
For the single-user approach, we mean that the relay node decodes the LDPC code $C_a$ (or $C_b$) from the received MAC signal by viewing the signal from node B (or node A) as the pure interference and then the XOR codeword based on the decoded codewords $\hat{\mathbf{c}}_a$ and $\hat{\mathbf{c}}_b$ is formulated for broadcasting.
To initiate the iterative decoding, it is of importance to compute the log-likelihood ratio (LLR) at any time epoch $k$ from the received signal $y_r$ (for brevity we omit the time epoch in what follows)
\begin{eqnarray}
  \label{eq:8}
  L_a &=& \log\frac{\Pr(c_a=1|y_r)}{\Pr(c_a=0|y_r)} \nonumber \\
      &=& \log\frac{\Pr(c_a=1,c_b=1|y_r)+\Pr(c_a=1,c_b=0|y_r)}{\Pr(c_a=0,c_b=1|y_r)+\Pr(c_a=0,c_b=0|y_r)}. \nonumber \\
      {}
\end{eqnarray}

Then, it is straightforward to show that
\begin{eqnarray}
  \label{eq:9}
  L_a &=& L_c y_r + \text{logcosh}\left(\frac{1}{2}L_c (y_r - 1)\right)  \nonumber \\
   &&- \text{logcosh}\left(\frac{1}{2}L_c (y_r + 1)\right),
\end{eqnarray}
where $L_c=\frac{2}{\sigma^2}$ and $\text{logcosh}(x)=\log(\cosh(x))$.

As shown in (\ref{eq:9}), there is no difference in computing the LLR output for nodes A and B. Hence, one can deduce that the single-user approach simply fails to work.

\subsection{Joint Network and LDPC Coding}

For joint network and LDPC coding, we consider the employment of the same LDPC code at both nodes A and B.
In this case, one have that $H_a=H_b=H$ and
\begin{equation}
 \label{eq:10}
    H (\mathbf{c}_a \oplus \mathbf{c}_b) = \mathbf{0}.
\end{equation}
Then, the relay tries to decode $\mathbf{c}_r=\mathbf{c}_a \oplus \mathbf{c}_b$ directly.
During the broadcast phase, the relay node transmits the XOR codeword $\mathbf{c}_r$ to both nodes A and B.
Then, nodes A and B decode $\mathbf{c}_r=\mathbf{c}_a \oplus \mathbf{c}_b$ based on the received signal vector and since they have $\mathbf{c}_a$ and
$\mathbf{c}_b$, they can obtain $\mathbf{c}_b$ and $\mathbf{c}_a$, respectively. Hence, the bottleneck is to decode $\mathbf{c}_r$ for the relay node during the MAC phase.

For the real MAC channel (\ref{eq:1}), we show, however, this difficulty can be solved smoothly.
Indeed, the MAC channel (\ref{eq:1}) can be equivalently viewed as
\begin{equation}
 \label{eq:11}
    y_r(k)=\psi(c_r(k))+w_r(k),
\end{equation}
where $\psi: \{0,1\} \rightarrow \{\pm 2, 0\}$ by abuse of notation. This equivalent channel is memoryless and is specified
by the conditional probability density function
\begin{eqnarray}
 \label{eq:12}
 p(y|x)= \left\{\begin{array}{c}\frac{1}{\sqrt{2\sigma^2}} e^{\frac{-y^2}{2\sigma^2}} \quad{}\quad{}\quad{}\quad{}\quad{}\quad{}\quad{}\quad{} \text{if}\quad{}x=1 \\
 \frac{1}{2} \frac{1}{\sqrt{2\sigma^2}}\left(e^{\frac{-(y-2)^2}{2\sigma^2}} + e^{\frac{-(y+2)^2}{2\sigma^2}}\right) \text{if}\quad{} x=0
 \end{array}\right.
\end{eqnarray}
This means that the equivalent MAC channel is typically an asymmetrical memoryless channel.

Assuming that the receiver has knowledge of the parameter $\sigma^2$. Then, if the optimal log-likelihood
detection is employed, the soft LLR output provided by the channel is given by
\begin{eqnarray}
  \label{eq:13}
  L_r &=& \log\frac{\Pr(c_r=1|y_r)}{\Pr(c_r=0|y_r)} \nonumber \\
      &=& \log\left[\frac{p(y_r|c_r=1)}{p(y_r|c_r=0)} \cdot \frac{\Pr(c_r=1)}{\Pr(c_r=0)} \right].
\end{eqnarray}

By noting (\ref{eq:12}), it follows that
\begin{eqnarray}
  \label{eq:14}
  L_r = \frac{2}{\sigma^2} - \text{logcosh}\left(\frac{2y_r}{\sigma^2}\right).
\end{eqnarray}

Once the soft LLRs referred to the XOR codeword $\mathbf{c}_r$ are computed, the relay then implements the standard iterative decoding over the Tanner-graph of $H$ for getting an estimate of $\mathbf{c}_r$.

\section{Joint Coding over Complex MAC Channel}
In practice, the real MAC channel model (\ref{eq:1}) is too ideal.
Hence, one often have to consider the following complex MAC channel
\begin{equation}
 \label{eq:15}
    \tilde{y}_r(k)=\tilde{h}_a x_a(k)+ \tilde{h}_b x_b(k)+z_r(k),
\end{equation}
where $\tilde{h}_a=|\tilde{h}_a|e^{j\theta_a}$ and $\tilde{h}_b=|\tilde{h}_b|e^{j\theta_b}$ are complex variables and $z_r(k)$ is the zero-mean complex additive Gaussian noise with variance of $E\{|z_r(k)|^2\}=2\sigma^2$.

If the channel is kept fixed during the transmission of a codeword, one can collect the signals to form a vector
\begin{equation}
 \label{eq:16}
     \tilde{\mathbf{y}}_r = \tilde{h}_a \mathbf{x}_a + \tilde{h}_b \mathbf{x}_b + \mathbf{z}_r,
\end{equation}
just as we encountered in (\ref{eq:2}).

Let $\Sigma =\{\xi_0,\xi_1\}$ be a discrete complex signal set, where $\xi_0 = \tilde{h}_a + \tilde{h}_b$ and $\xi_1 = \tilde{h}_a - \tilde{h}_b$.
With the same idea developed in Section-III, one can show that an equivalent channel model for the XOR codeword $\mathbf{c}_r=\mathbf{c}_a \oplus \mathbf{c}_b$
takes the form of
\begin{equation}
 \label{eq:17}
    \tilde{y}_r(k)=\psi(c_r(k))+z_r(k),
\end{equation}
where $\psi: \{0,1\} \rightarrow \{\pm \xi_0, \pm \xi_1\}$. This equivalent channel is memoryless and is specified
by the conditional probability density function
\begin{eqnarray}
 \label{eq:18}
 p(\tilde{y}|x)= \left\{\begin{array}{c}\frac{1}{2}\frac{1}{2\sigma^2} \left(e^{\frac{-|\tilde{y}-\xi_1|^2}{2\sigma^2}} + e^{\frac{-|\tilde{y}+\xi_1|^2}{2\sigma^2}}\right) \text{if}\quad{}x=1 \\
 \frac{1}{2} \frac{1}{2\sigma^2}\left(e^{\frac{-|\tilde{y}-\xi_0|^2}{2\sigma^2}} + e^{\frac{-|\tilde{y}+\xi_0|^2}{2\sigma^2}}\right) \text{if}\quad{} x=0
 \end{array}\right.
\end{eqnarray}
This means that the equivalent MAC channel is typically an asymmetrical memoryless channel.
The LLR output can be calculated as
\begin{eqnarray}
  \label{eq:19}
  L_r &=& \text{logcosh}\left(\frac{\Re[\tilde{y}_r \xi_1^*]}{\sigma^2}\right) - \text{logcosh}\left(\frac{\Re[\tilde{y}_r \xi_0^*]}{\sigma^2}\right) \nonumber \\
     && - \frac{|\xi_1|^2-|\xi_0|^2}{2\sigma^2}.
\end{eqnarray}

For the normalized equal-power case, namely, $|\tilde{h}_a|=|\tilde{h}_b|=1$, this complex MAC channel is often characterized by the carrier-offset $\Delta \theta = \theta_b-\theta_a$.

\section{Simulation Results}

\begin{figure}[htb]
   \centering
   \includegraphics[width=0.45\textwidth]{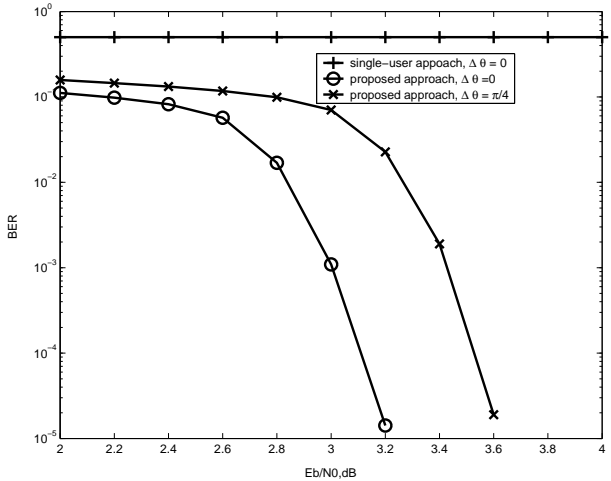}
   \caption{Performance comparison between the single-user approach and the proposed approach.}
   \label{fig:1}
\end{figure}

As the bottleneck of the bi-directional relaying system lies in the processing capability of the relay node and its performance.
For the proposed joint network and LDPC coding scheme, it is the duty of the relay to reproduce the XOR codeword $\mathbf{c}_r=\mathbf{c}_a \oplus \mathbf{c}_b$.
Hence, we mainly focus on the performance of the XOR codeword $\mathbf{c}_r$. The performance is closely related to the energy per bit and the received noise variance.
For a simple comparison with the standard BPSK modulated AWGN channel, we still assume that  $\sigma^2=(2 R_c E_b/N_0)^{-1}$ for the real MAC channel model of (\ref{eq:1}), where $E_b$ denotes the energy per bit for node A (or B).
This is possible as we assume that both nodes A and B employ the same LDPC code with $R_a=R_b=R_c$ and the energy per bit for both nodes A and B keeps the same.

The considered code is a (3,6)-regular LDPC code with codewords of length $N=4096$, which is constructed by a progressive-edge-growth
algorithm reported in \cite{Hu_PEG}. A maximum number of demodulation iterations is $60$.
For decoding of LDPC codes, the normalized min-sum algorithm (NMSA)\cite{ChenNMSA} is assumed, which can speed the simulations with little degradation on the final performance. The scaling factor for the NMSA decoding is set to $0.85$.

The BER performance is shown in Fig. \ref{fig:1} for the real MAC channel. Also included is the normalized equal-power complex MAC channel with $\Delta \theta= \pi/4$. As shown, the joint network and LDPC coding approach performs well while the single-user approach simple fails to work as predicted previously.

\section{Conclusion and Future Work}

We have presented a joint network and LDPC coding scheme for bi-directional relaying.
It is found that the multiple-access channel at the relay node can be viewed as an equivalent asymmetrical channel where the channel input codeword is the XOR codeword between nodes A and B. In simulations, we employ a (3,6)-regular LDPC code, which is far from the optimality for the equivalent asymmetrical channel. Hence, it is of interest to consider the design of LDPC codes for the asymmetrical MAC channel by employing the idea of \cite{Wang_Asym}.


\end{document}